\documentclass[10pt]{bmc_article}    

\usepackage[pdftex]{graphicx}
\usepackage{subfig}
\usepackage{cite} 
\usepackage{url}  
\usepackage{ifthen}  
\usepackage{multicol}   

\usepackage[latin9]{inputenc}

\newboolean{publ}

\newenvironment{bmcformat}{\begin{raggedright}\baselineskip20pt\sloppy\setboolean{publ}{false}}{\end{raggedright}\baselineskip20pt\sloppy}

\begin{document}
\begin{bmcformat}


\title{TinkerCell: modular CAD tool for synthetic biology}
 

\author{Deepak Chandran\correspondingauthor$^{1}$
       \email{Deepak Chandran\correspondingauthor - deepakc@u.washington.edu}%
      \and
         Frank T. Bergmann$^{1,2}$
       and 
         Herbert M. Sauro$^{1}$
      }


\address{%
    \iid(1)Department of Bioengineering, University of Washington, Box 355061,%
        William H. Foege Building, Room N210E, \\
        Seattle, WA, USA 98195-5061 \\
    \iid(2)Keck Graduate Institute, %
           535 Watson Drive, Claremont, CA, USA, 91711
}%

\maketitle


\begin{abstract}
\paragraph*{Background: } Synthetic biology brings together concepts and techniques from engineering and biology. In this field, computer-aided design (CAD) is necessary in order to bridge the gap between computational modeling and biological data. Using a CAD application, it would be possible to construct models using available biological ``parts" and directly generate the DNA sequence that represents the model, thus increasing the efficiency of design and construction of synthetic networks.
      
\paragraph*{Results:} An application named TinkerCell has been created in order to serve as a CAD tool for synthetic biology. TinkerCell is a visual modeling tool that supports a hierarchy of biological parts. Each part in this hierarchy consists of a set of attributes that define the part, such as sequence or rate constants. Models that are constructed using these parts can be analyzed using various C and Python programs that are hosted by TinkerCell via an extensive C and Python API. TinkerCell supports the notion of a module, which are networks with interfaces. Such modules can be connected to each other, forming larger modular networks. TinkerCell is a free and open-source project under the Berkeley Software Distribution license. Downloads, documentation, and tutorials are available at www.tinkercell.com.

\paragraph*{Conclusions:} The final goal of a CAD application is to make the design and construction process easier. The biggest limitation in synthetic biology at present is the limited knowledge on the precise kinetics and parameters that explain the dynamics of biological parts. TinkerCell is built under the assumption that this limitation will be resolved, allowing models to represent the real system more accurately. TinkerCell can assist in pushing the field toward this goal by providing a framework for constructing models using parts and modules. Because TinkerCell associates parameters and equations in a model with their respective part, parts can be loaded from databases along with their parameters and rate equations. The modular network design can be used to exchange modules as well as test the concept of modularity in biological systems. The flexible modeling framework along with the C and Python API allows TinkerCell to serve as a host to numerous third-party algorithms.
\end{abstract}

\ifthenelse{\boolean{publ}}{\begin{multicols}{2}}{}

\section*{Background}

Systems level modeling has lead to a clearer mechanistic understanding of biological systems and provided mathematical explanations for experimental observations \cite{hasty2001dgn,isaacs2003pam,ozbudak2004mlu}. One of the consequences of such understanding has been the ability to design and engineer synthetic biological networks in microorganisms, which has given rise to the field of Synthetic Biology \cite{endy2005fe}. The idea of engineering biological systems brings together concepts from various fields. While the laboratory procedure at present is borrowed from genetic engineering, concepts such as abstraction and interchangeable parts are taken from computer science and electrical engineering. Synthetic biology introduces the notion of biological ``parts" \cite{endy2005fe}, which are individual components that can be assembled in different ways to construct synthetic networks with different functions. Networks built by different engineers can then be reused to construct larger networks, much like a programmer using existing subroutines to build a new program more efficiently. Since synthetic biology is a young field, the best practices for making molecular biology interchangeable and programmable have not been established. Nonetheless, various success stories \cite{elowitz2000so,gardner2000cgt,entus2007dai} have shed light on the enormous potential of synthetic biology to understand fundamental science \cite{endy2005fe} or create new solutions for applications ranging from fuels \cite{lee2008mem} to medicine \cite{anderson2006ec}.

The terminology, software, and laboratory procedures required to push synthetic biology forward are still in development. However, one can anticipate that certain key concepts such as standarization and modularity will become commonplace in synthetic biology \cite{heinemann2006sbp}. In synthetic biology, the need for standardization exists at several levels. At one level, there is a standard for defining parts. While biological parts are the individual components for building a network, \textit{standardized} parts contain additional restrictions that are intended to make synthetic networks easier to build, more reliable, and easier to exchange. An existing standard is the standard assembly \cite{shetty2008eb}, which has made DNA assembly simpler. In future, it is anticipated that standards will also exist for describing the dynamics of a part; for example, standard promoter parts might contain a ``strength" value, describing its efficiency in recruiting RNA polymerase under some standard environmental condition \cite{kelly2009ma}. This leads to the second level of standards, which describes parts in a computer-readable format such as the Resource Definition Language \cite{doi:1721.1/45537}, so that searching and organizing parts can be automated. Under such a framework, parts could be organized by a defined ontology. The third level of standards applies to computational models. While there are existing standard formats for representing biological models \cite{hucka2003sbm,lloyd2004ci}, synthetic biology models might require additional information such as the DNA sequence or specific information about the parts that are needed to physically construct the network.

Several issues in modeling for synthetic biology have to be addressed. One is that the models need to support kinetic information as well as information for constructing the network. Second, due to the fact that synthetic biology is a merger of multiple fields, a large number of analyses may be possible on synthetic networks, ranging from dynamical systems analysis to analysis of the DNA sequence or statistics on the part usage in the model. The software application presented in this work, TinkerCell, will address these issues. While TinkerCell targets synthetic biology, it addresses several issues that are equally applicable outside the field. These issues include support for third-party libraries, modular design of networks, flexible modeling framework, and flexible visual formats. 

TinkerCell is an improvement on a similar effort, Athena, by the same authors \cite{chandran2009amc}. Athena addressed issues such as modular design of networks and biological parts. However, the underlying design of Athena was not as flexible as it was intended to be. For example, Athena was not able to support an ontology of biological parts. Due to such limitations in the design, the project was completely restarted under a different name.

\textbf{Existing computational tools : }
Numerous software tools exist that allow construction and analyses of models using scripts or visual interfaces. A comprehensive list can be found at www.sbml.org. Some of these applications include Jarnac \cite{sauro1993sgp} , JDesigner \cite{sauro2003ng} , CellDesigner \cite{funahashi2003cp} , BioTapestry \cite{levine2005grn} , PySCeS \cite{olivier2005mcs}, BioJADE \cite{goler2004bd}, little-B \cite{mallavarapu2009pmm}, SynBioSS \cite{hill2008ssb}, ProMot \cite{ginkel2003mmc, marchisio2008cds,mirschel2009pmm}, and Antimony \cite{antimony}. Each of the applications that have their respective advantages. For example, Jarnac and PySCeS are highly flexible due to their programming interface. Little-B is similar, building on Lisp, but it supports modules in addition. CellDesigner offers a plug-in interface, which has permitted the community to add new features. BioTapestry has a simple visual depiction for genetic networks and genetic modules. BioJADE and SynBioSS, being a synthetic biology applications, supports a parts database \cite{registryofparts}. Antimony is a library defining a text-based language for constructing biological modules and composing models using those modules. ProMot is similar but also has a visual interface that supports modularity as well as using parts to compose a circuit \cite{marchisio2008cds}. We considered that TinkerCell should, at the least, contain the qualities of each, which include the flexibility of programming, plug-in support, database support, modularity, and ability to construct genetic networks. However, the intent behind TinkerCell is not to create yet another modeling application, but to create an application that can serve as a host to the algorithms, data, and ontologies that the synthetic biology and the systems biology community can offer. Unlike most modeling applications, TinkerCell does not impose a particular modeling method (e.g. differential equations), visual representation, or a strict definition of a model.

\section*{Results and Discussion}

The application named TinkerCell is a synthetic biology CAD tool for visually constructing and analyzing biological networks. The visual interface itself does not have any analysis capabilities. The analyses are performed by third-party C or C++ libraries and Python modules that interface with TinkerCell. Therefore, TinkerCell can be thought of as a front-end to C and Python programs. The visual interface of TinkerCell must remain as open as possible so that it can host a variety of third-party programs. This is achieved by remaining oblivious to the \textit{meaning} of a model. The meaning is given by the C and Python programs that utilize the information stored in the model. This framework allows TinkerCell to remain open to different modeling methods used by different C or Python programs (see Figure \ref{fig:Structure1}).

The visual side of TinkerCell provides the interface for drawing nodes and connections as well as constructing modules and compartments (see Figure \ref{fig:compartment_module}). Nodes represent physically distinct biological components such as proteins or metabolites as well as regions of the DNA, such as promoter or coding region. Connections refer to reactions such as catalysis, transcription, or translation, as well as other relationships between nodes such as relative location on the DNA, i.e. upstream or downstream. Compartments represent spatially separated regions such as cells or sub-cellular compartments. Modules are networks with interfaces that can be connected with each other to form larger models. The visual appearance is similar to cartoon diagrams that are used in text-book diagrams to explain pathways (see Figure \ref{fig:threecells}).

Within each part and connection, TinkerCell stores information pertaining to that component of the model. A computational model is typically described by the variables in the system, the dynamics of the system (e.g. differential equations), and the parameters and other functions used in the model; TinkerCell takes a slightly different approach: the parameters and functions are stored within the components of the model. For example, a traditional model of gene regulation might have a rate equation for transcription, translation, and protein degradation that all use parameters defined in the model. In TinkerCell, the rate equations belong with the transcription regulation connection, and parameters such as the promoter efficiency belong with the parts used in the model. This allows a user to replace one of the parts in the model, which would automatically result in changes to the rate equations or parameters. In a traditional model setting, one would need to identify which parameters are associated with each of the replaced parts and change them individually. Other information such as DNA sequence, annotation, authorship, or experimental data are also stored within the components in the model. 

TinkerCell serves as a drawing and information storage application; the task of analyzing the model is handed to C and Python programs. For this reason, TinkerCell provides an extensive C and Python Application Programming Interface (API), which allows third-party programs to access information in TinkerCell models and perform various analyses. The API also provides a rich set of output functions, allowing C and Python functions to fully integrate with TinkerCell's visual interface. The motivation for the separation of the visual interface and functionality is to allow users with programming experience to add new functionality to TinkerCell. These new functions can then be shared among other users. Further, existing C libraries or Python modules can also be integrated into TinkerCell. Thus, TinkerCell serves as a host for programs that the community can contribute. Currently, it hosts a handful of C libraries such as simulation and linear programming libraries, and a few Python modules, such as SciPy \cite{jones2001sos}, PySCeS \cite{olivier2005mcs}, and NerworkX \cite{hagberg2004nld}.

TinkerCell is designed to meet future standards in synthetic biology. We anticipate that one of the results of standardization efforts in synthetic biology will be an organized ontology describing biological parts and reactions.  
The set of parts and reactions available in TinkerCell are loaded from two Extensible Markup Language (XML) files, thus keeping TinkerCell flexible so that it can cope with future standards regarding part and connection ontology (see Figure \ref{fig:overview_window}). 

In summary, there are three features by which TinkerCell is uniquely suitable as a CAD tool for synthetic biology. One, TinkerCell can construct models from a catalog of parts that are obtained from an outside source. The model dynamics are determined by the parts used in the model rather than parameters of the model, as in traditional modeling. Second, networks can be modularized, allowing one user to build new designs by utilizing existing ones. Being able to build upon the constructs from previous work is essential for any engineering discipline. Third, TinkerCell can support a large number of functionality because it is not restrictive to the modeling approach and has an extensive C and Python API, which has been demonstrated to support functions ranging from simulations to sequence analysis. In addition to these primary features, TinkerCell's architecture and visual framework are kept as flexible and open as possible in order to welcome contributions from the community.

\label{architecture}
\section*{Methods}

TinkerCell is organized in layers (see Figure \ref{fig:Structure2}), with the bottom-most layer being a generic API for constructing network drawing programs and the top-most layer being a scripting interface. TinkerCell is extensible at each of the layers. There is an additional level of flexibility provided by the XML files that load the ontology of parts and connections (see section \ref{ontology}). The goal of this structure is to foster contributions from the community. 

\subsection*{The Core library}
The TinkerCell ``Core" library is a C++ library built using the Qt Toolkit 4.5.0 \cite{toolkit2007tws}. The Core library contains the data structures and functions for drawing parts and connections and storing information associated with those items (see Figure \ref{fig:Structure1}). It provides an API which can be used to build new network drawing programs. The Core library also provides support for C functions and a generic console window, which is currently used as a Python console but can later be used to support other languages as well. The complete TinkerCell application is a collection of plug-ins that utilize the TinkerCell Core library in order to provide the full set of features. Throughout this article, the name ``TinkerCell" will refer to the collection of plug-ins and the TinkerCell Core library, because the Core library alone has no modeling capability. Documentation on the Core library is available at www.tinkerell.com.

\subsection*{Plug-ins and flexible modeling framework}

TinkerCell uses the notion of plug-ins. Plug-ins are C++ programs that add new features to TinkerCell without altering the existing code (see Figure \ref{fig:Structure1}), thus allowing programmers to extend TinkerCell. The current set of plug-ins are responsible for storing parameters, rate equations, and other information relevant to modeling as well as storing sequence and other details needed for physical construction of the network. The plug-ins also provide the user interface for editing these values. The complete TinkerCell application is composed of a collection of plug-ins that build on the TinkerCell Core library (see Figure \ref{fig:Structure1}). Plug-ins are responsible for providing visual features such as scaling and coloring as well as modeling features such as loading the catalog of parts and connections and inserting parts and connection from that catalog. See section \ref{divisionOfLabor} for details on how the tasks are divided between the Core library and the plug-ins. 

An important role of plug-ins is to determine what information is stored by a TinkerCell model. For example, the Numerical Attributes plug-in and the Reaction Rates plug-in, together, ascertain that a TinkerCell model will contain sufficient information to generate the stoichiometry matrix and rate equations for the model (see Figure \ref{fig:Attributes_and_rates_example}), which are required to generate differential equation models or stochastic models. Additional information, such as function definitions and events, are added by other plug-ins. While the existing plug-ins focus on building models based on stoichiometry and rates, new plug-ins can focus on other modeling approaches, such as rule-based or Boolean modeling. The current set of plug-ins do not include such approaches, but the underlying structure of TinkerCell is not limited to any one modeling approach.

\subsection*{C and Python interface}

Plug-ins that are related to modeling are almost always supported by C or Python functions. For example, the Numerical Attributes plug-in and the Reaction Rates plug-in enrich the model with sufficient information to generate dynamic models, but the plug-in itself does not perform any analysis. Instead, each plug-in provides an API containing functions such as ``getParameter", ``getRate", or ``getStoichiometry", allowing third-party C and Python programs to obtain the necessary information needed to carry out a simulation or other analyses. Almost every plug-in in TinkerCell exposes an API, creating a rich C and Python API with over a hundred different functions. The collection of API functions allow C and Python programs to get or set kinetic information such as parameters, rates, functions, and stoichiometries, or change visual aspects such as positions, colors, and line widths (see Figure \ref{fig:Visual_C_outputs}). The API also provides ways for C and Python programs to bring up dialogs, asking the user for inputs (see Figure \ref{fig:createInputWindow}).

All C and Python programs automatically run on separate threads, allowing a user to continue working on a model while a time consuming task is running. TinkerCell's default simulators compile the model as a separate C program, allowing the simulations to run at the speed of a compiled C program. This strategy is used by the deterministic simulator as well as the stochastic simulator, where the speed gain is more visible.

The C API has been used to include the following C programs in TinkerCell: deterministic simulation and steady state analysis using the Sundials library \cite{hindmarsh2005ssn}, stochastic simulation using a custom C library, flux balance analysis using LPsolve \cite{berkelaar:los}, and some custom C programs. The Python API has been used to include the SciPy module \cite{jones2001sos}, PySCeS module \cite{olivier2005mcs}, NerworkX  module \cite{hagberg2004nld}, a custom programs for generating transcriptional rate formulas, and another custom program for generating a FASTA file with the DNA sequences of the parts in the model.

A user may add new C functions to TinkerCell by placing the C dynamic library files in the TinkerCell's ``Plugins/c" folder and new Python functions by adding the name of their script files in ``pythonscripts.txt". The C programs can appear as buttons or menu items, and the Python programs will automatically appear as buttons in the TinkerCell application, making integration of new function simple.

\label{modular_design}
\subsection*{Modular design of networks}
In TinkerCell, new models can be constructed by connecting existing networks to one another. Such composable networks, or modules, are constructed by defining interfaces. A user may declare one or more components of a module as interfaces. These interfaces can then be connected with one another, which indicates that the respective components of the modules have been merged. For example, if a user wishes to construct a cascade of phosphorylation cycles \cite{sauro1993bng}, then they may simply connect the kinase of one module to the phosphorylated protein in the other module, which will merge the two. Merging two molecular species means that the two visually presented species represent the same physical entity (see Figure \ref{fig:KinaseModule}). The connections do not alter the modules themselves. 

\subsubsection*{Genetic modules}
Another way to connect two or more modules is by introducing new reactions between the interfaces. The synthetic biology community is familiar with the notion of modules in the context of DNA. For example, one module might be responsible for regulating a promoter, and the next module might be situated downstream of the promoter, thus implying a regulatory relationship between the two modules. In TinkerCell, this is achieved by connecting the promoter of the upstream module to the gene in the downstream module  (see Figure \ref{fig:PoPS}). This connection belongs to a special family of connections that represents RNA polymerase activity along a DNA strand, known as ``PoPS" (Polymerase Per Second \cite{kelly2009ma}) in the synthetic biology community.

\subsubsection*{Modules as composite parts}
The synthetic biology community is familiar with the notion of a ``composite part", or a biological part that has been constructed by assembling other parts together \cite{registryofparts}. TinkerCell can represent such an object as a module (see Figure \ref{fig:composite_part}). Each sub-part inside the composite part retains its original set of attributes and other information while the composite part can have its own family and set of attributes. The C and Python API can be used to access the parameters, annotation, authorship, and other identification numbers for each of the sub-parts as well as the composite part.

\subsection*{Support for standards}

It is our anticipation that the growth in synthetic biology will result in a general agreement in the community about what attributes are sufficient to define a biological part. For example, it might be sufficient to state the sequence, binding sites, and RNA polymerase binding affinity to define a promoter part; there will be a similar list of sufficient attributes for other biological parts. Further, ontologies or other ways of organizing biological parts will become necessary when annotating a model. TinkerCell is structured to support such an organized set of parts with attributes. Every item in TinkerCell belongs to a ``family". The families themselves are defined in an XML file and are organized as a hierarchy. For example, the family called ``Transcription Factors" belongs under the family named ``Proteins". Each family is defined with a set of attributes. For example, genes contain a ``sequence" attribute. In addition to sequence, promoters or Ribosomal binding sites contain a ``strength" attribute, which is a number describing the relative strength of the promoter or RBS. The families and their set of attributes are not intended to be defined by TinkerCell; the hierarchy of parts is intended to be obtained from a database of biological parts in future. Just as every part in TinkerCell model is identified with a family, so are connections defined with a family. Examples of connection families are ``Biochemical reaction", ``Binding reaction", and ``Transcriptional regulation". Each connection family also contains attributes; for example, ``Transcription regulation" contains parameters named ``h", the Hill coefficient, and ``Kd", the dissociation constant. The connection families are also loaded from an XML file, hence they are not intended to be defined by TinkerCell either.

\label{ontology}
\subsubsection*{Using ontology in models}

The advantage of constructing models that also define the family of each item is that plug-ins or third-party programs can utilize the fact that certain parts or connections will always have certain parameters. An example is the ``Hill equations" Python script, which automatically generates rate equation using the fractional saturation model \cite{shea1985oc}. The script utilizes that fact that every ``Transcription regulation" connection contains a ``Kd" and a ``Hill" parameter, and it uses these parameters to automatically generate the transcription rate equation. Another Python script included with TinkerCell searches the RegulonDB database \cite{gamacastro2007rvg} for ribosomal binding sites, promoters, and terminators from \textit{E.coli} and automatically fill in appropriate attributes based on the type of part that is selected; for example, when the script sees a promoter, it adds the sigma factor information to the part, and when it sees a terminator, it adds information about whether or not it is a rho-dependent terminator. An example plug-in that uses the family information is the DNA sequence viewer (see Figure \ref{fig:SequenceView}), which assumes that every item of the family ``DNA" will contain an attribute called ``sequence" and is able to present the information visually. One can imagine other functions where features of the promoter, such as the sigma-factor, may be used within the model to provide additional details, and the user can load different promoters from a catalog of biological parts to test how the different promoters affect the network. 

It should be noted that the families and their attributes are not defined within TinkerCell, so the algorithms are general for any future ontology that the synthetic biology community will adopt. In addition to the family information, each part and connection in TinkerCell contains its own annotation, storing information such as authorship, references, and date. As with any other information in TinkerCell, all of this information is accessible and editable from C and Python. 

\label{standard_visual_format}
\subsubsection*{Standard visual formats}
Standard visual formats, such as the standard symbols that are used to draw electronic circuits, allow network diagrams to be unambiguous. While systems biology has made progress to standardize visual representations of biological systems \cite{kohn2006cdb}, such standards for synthetic biology are still in development \cite{BOGOL}. TinkerCell has been designed so that it will be able to cope with a variety of visual standards or even multiple visual standards. This is achieved by having a flexible visual representation. The visual representation of proteins, genes, promoters, and the rest of the parts are stored as XML files. These files are generated using a polygon drawing program that comes with TinkerCell (see Figure \ref{fig:replace_graphics}). The same file format is used for arrow heads and decorators such as phosphorylation sites, which keeps visual representations for those objects flexible as well.

In order to cope with a future visual standards, new files will be generated using the polygon drawing program that match the prescribed visual standard. 

\subsection*{Genetic networks}

Since the majority of the current synthetic networks are genetic networks, TinkerCell gives special consideration to them. TinkerCell provides three different ways for modeling genetic networks, allowing the modeler to choose which is most suitable. 

\subsubsection*{Fractional saturation models : } The first method of modeling gene regulatory networks is using the equilibrium assumption for transcription factor binding \cite{shea1985oc}. Under this model, a single rate expression is generated that captures the probability of the promoter being in an active state. This rate expression assumes that the transcription factor binding and unbinding are at equlibrium. TinkerCell automatically generates the rate expressions when a user connects transcription factors to a gene. See Figure \ref{fig:grn_simple}.

\label{geneticnetworks}
\subsubsection*{Using parts to model genetic networks: } The second method of modeling gene regulatory networks involves the same equilibrium assumption mentioned above. However, the entire gene is not represented as a single item; it is split into a set of distinct parts (see Figure \ref{fig:grn_parts}). Each part can be moved individually. The rate expression can be described in terms of the promoter strength and ribosomal binding site strength, thus allowing a user to swap one part with another, e.g. replace a weak promoter with a strong one. The rate expression remains defined in terms of the parts that are directly upstream of the coding region. See Figure \ref{fig:grn_parts}.

\subsubsection*{Explicitly defining intermediate steps: } A more elaborate way of describing gene regulatory networks is to define each reaction in the transcription and translation process, including the movement of RNA polymerase across the gene \cite{kosuri2007ts}. Creating such a model would be nearly impossible without some sort of automation, and TinkerCell provides a function for automating this process (see section \ref{automated kinetics}). While such models can avoid the use of equilibrium assumptions, simulation can be time-consuming due to the fact that each transcription and translation step is comprised of multiple reactions. See Figure \ref{fig:grn_full}.

\subsection*{List of functions provided through the C and Python interface}

While the visual side of TinkerCell provides the interface for constructing a model, the analyses are carried out by C and Python functions. New C and Python functions may be added to TinkerCell by listing the program names in a text file. Programs that are currently listed in those files provide various functionalities, as described below.

\subsubsection*{Fast simulators}
The default simulators use existing C libraries that perform deterministic and stochastic simulation. For example, the Gillespie simulator uses a custom Gillespie algorithm C library available from tinkercell.googlecode.com. The C library accepts a user-defined stoichiometry matrix and propensity function. The Gillespie simulator in TinkerCell takes the stoichiometry matrix and reaction rates from the TinkerCell model and generates the C code containing the stoichiometry matrix and propensity function. This C code is compiled and linked against the existing Gillespie algorithm library. The compiled C program performs the simulation and outputs the result to TinkerCell's plot window. Therefore, the simulation itself is performed by a compiled C program, which means that the simulation will be fast. Conventional simulators interpret the rate equation string and construct an internal representation of the equation, which is used to compute the rate value during every iteration; this is generally an order or magnitude slower than the compiled program. The same idea is used by the deterministic simulation, which uses the Sundials CVODE numerical integrator \cite{hindmarsh2005ssn} to perform the simulation. One drawback is that there is a short time delay when compiling the code, which may be more than the time for simulation for simple networks. Therefore, the speed gain due to this strategy is only visible for larger networks. However, this demonstrates how existing C libraries can be used without modification to create new simulators for TinkerCell. The Tiny C Compiler \cite{bellard:ttc} and GNU C Compiler \cite{GCC} are used to perform the compilation in windows and unix, respectively. 

Further, the simulation programs use an function called ``createInputWindow" that is provided by the API. This function allows C programs to create simple input windows where the user can input parameters such as the time for simulation or the type of simulation (see Figure \ref{fig:createInputWindow}). There are several such functions that allow C and Python programs to interact with the user through visual interfaces.

\subsubsection*{Visual inputs and outputs for C and Python programs}
While C and Python programs are generally expected to take command-line inputs and produce command-line outputs, TinkerCell provides functions that allow C and Python functions to have more interactive inputs and outputs. TinkerCell's C and Python API contain functions for displaying dialogs or input tables. The outputs from these dialog and tables are returned to the calling C or Python program. The C or Python program can then use the information from the model and the inputs from the user to perform some analyses. As an output, the C and Python programs can change the size or color of items in the model, circle items in the model, or display numbers or strings next to items in the model. At the same time, the programs may also print to the TinkerCell command-line window. 

\textbf{Flux balance analysis :   }
The flux balance analysis is one example where a C program produces visual outputs. This particular program uses a C++ plug-in to create a custom input window. The C++ plug-in sets up the windows, tables, and buttons specifically for flux balance analysis. This window simply serves as a user interface to a C program. The C program that performs flux balance analysis is less than 100 lines long. It receives input from the C++ plug-in and uses the LPSolve C library \cite{berkelaar:los} to do the optimization. The C program then changes the width of the reaction arcs according to the result from LPSolve. The optimal fluxes are also displayed next to the reactions arcs. 

\textbf{Sensitivity analysis :   }
The sensitivity analysis is yet another example of visual output (see Figure \ref{fig:Visual_C_outputs}). The senstivity converts the TinkerCell model into a PySCeS \cite{olivier2005mcs} model and uses PySCeS to perform the analysis. The script then takes the PySCeS output and colors the reactions in the network according to their control coefficients.

\textbf{Accessing \textit{E.coli} genetic parts through RegulonDB :   }
RegulonDB \cite{gamacastro2007rvg} houses a large set of promoters, transcription factor binding sites, and ribosomal binding sites for the \textit{E. coli} K12 strain. Through a Python script, a user can load sequences and other information such as promoter type into TinkerCell models. The script provides the user with a visual interface for selecting the parts. The script also limits the search by using known information. For example, if the transcription factor LacI is repressing a promoter, then the script will show only those binding sites that are targeted by LacI (see Figure \ref{fig:regulonLoad}), allowing a user to automatically generate the correct network using real parts. The Python script is able to use the family information of objects in the model to identify whether it is a promoter, RBS, coding region, or transcription factor. It would be impossible to search for an appropriate part with this information in the model, which demonstrates how the meta-data is useful for database search.

\label{automated kinetics}
\subsubsection*{Automated kinetics}
One of TinkerCell's objectives is to combine the appeal of visual network design with the flexibility of programming. Automatic generation of kinetics is a good example where this objective is met. There are three functions currently included with TinkerCell that fall in this category. First, the ``Hill equation" Python function automatically generates the fractional saturation model for transcription regulation based on the activators and repressors of a promoter. Second, the ``binding reactions" C program automatically generates all the intermediate stages of a protein that form multiple complexes. Third, the ``multiple step process" function can automatically insert intermediate steps into any given reaction. For example, if the conversion of one molecule to another requires several intermediate stages, this function can be used to automatically generate those intermediates reactions. Gene regulatory networks are one area where this function can have addition relevance.

\textbf{Automatically generate multiple conformations of a protein:   }
The automatic generation of kinetics is a good example of how this objective is met. Consider a simple situation of two small proteins binding to a larger protein. In addition to the three proteins, there are three other complexes. However, the number of possible complexes grows exponentially with the number of binding proteins, thus modeling every complex explicitly becomes impractical. Generally, a modeller may choose to ignore such details due to the difficulty in modeling. TinkerCell offers a feature for automatically generating all necessary combinations (see Figure \ref{fig:binding_kinetics}). This feature is provided by a C function that recursively generates all the intermediate states and reactions. Integrating the existing function into TinkerCell required adding code that obtained ``Binding reactions" in the set of items that the user has selected and passing the stoichiometry for those reactions to the existing C function. The output from the C function is then used to modify the TinkerCell model.

\textbf{Automatically generate intermediate stages of a multi-step process:   }
Another automated feature is converting a single reaction into a multi-step reaction. This feature was inspired by the fact that transcription can be understood as several individual reactions, where each reaction is catalyzed by RNA polymerase. Modeling genetic networks in this manner can have significantly different outcomes in some situations \cite{kosuri2007ts}. It would be impossible for a user to draw hundred of reactions for each transcriptional process in a model. TinkerCell offers an automated function, again written in C, that will convert one reaction into multiple reactions. The visual model remains unchanged, but the selected reactions will represent multiple reactions even though only one will be visible (see Figure \ref{fig:tabasco_kinetics}). While an alternative approach might be delayed differential equations, this approach allows other details such as leaks in transcription and different delays based of the type of process.

\subsection*{TinkerCell file format}

TinkerCell stores the entire model as a single XML file, which contains both the model information as well as the graphical information. All the graphical objects in TinkerCell are generated from XML files, and therefore they can be saved as XML files. The model itself is stored as a set of components and information associated with those components. The information associated with each component is a set of tables containing the parameters, reaction rates, function definitions, events, DNA sequence, annotation, and any other information pertaining to that component in the model. The file format is therefore just a list of objects, their family information, and data tables associated with the objects. The datatables are labeled as ``parameters", ``reaction rates", ``text attributes", and so on. The format is generic, which reflects TinkerCell's underlying structure.

\section*{Conclusion}

We have described an application that combines the flexibility of programming with a visual interface as well as provides the structure for supporting standards and exchange of information. The intent behind TinkerCell is not to create yet another modeling application. TinkerCell is an application for bringing together models, information, and algorithms. Thus, it serves a host by itself. The C and Python interface have been demonstrated to be very versatile because simulation libraries as well as linear programming libraries were easy to integrate. Similarly, various Python modules have been readily integrated as well. Therefore, TinkerCell has the features to serve as a hub that can allow programmers, experimentalists, modellers, and theoreticians to exchange models and information. 

The modular modeling framework of TinkerCell can serve two purposes. The first is to aid in the engineering process by allowing one modeler to use existing modules to design new networks. The second purpose is that modeling with modules can be used to explore the possible implication of functional modules in biology. It is an open question of whether a module's functionality is retained when it is placed in different situations \cite{delvecchio2008mc}. There may be common themes behind modules where functionality is retained. Answers to such questions will be highly relevant to synthetic biology, since the ability to construct one system using another is a central theme in engineering. TinkerCell can serve as a platform for testing how different modules behave and what types of modules are able to retain their functional identity. The flexible modeling framework provides such experiments to be carried out at different levels of detail. For example, some issues regarding modularity may not be visible when using fractional saturation models of gene regulation, but they may be evident when the reactions are modeled explicitly. Since TinkerCell provides different ways of modeling, it is suitable for studying such questions.

\subsection*{Future work}
Efforts will be made in order to assure that the interface for constructing models as well as integrating new programs is suitable for different types of users. For example, an expert in the field of modeling might wish to see and edit all the details of the model, whereas another user may desire automated features. Similarly, users may wish to contribute programs that may perform mathematical analyses or exchange information with a database. TinkerCell is designed to meet all of these demands, but user inputs will decide what changes need to be made.

Models in TinkerCell are constructed visually. While text-based modeling is supported through the Antimony plug-in, it does not provide all the features that are available through visual model construction. At present, we are adding a text-based model construction framework that will contain all the features of the visual modeling framework. The syntax for the text-based framework will be kept flexible. With the completion of this feature, users can build models using text-based and graphical approaches.

In addition to improving the interface, TinkerCell will remain active in standardization efforts for synthetic biology, which may include standard ontologies or visual standards. While conversion to the \textit{de facto} standard file format in systems biology, the Systems Biology Markup Language (SBML) \cite{hucka2003sbm}, is available in TinkerCell through Antimony or PySCeS, this conversion will be improved in the near future by making this a direct conversion as well as exporting and importing layout information.

The modular approach to modeling will be explored further by using it to construct small modules and investigating the types of systems that can emerge by connecting simple modules in different ways. Other efforts will be directed toward adding interfacing existing C and Python programs with TinkerCell in order to add additional functionalities such as optimization or different types of simulations. 

\section*{Acknowledgements}

This work was partly funded by the National Science Foundation (Id 0527023- FIBR) and Microsoft's Computational Challenges in Synthetic Biology 2006 Award.

\pagebreak

\section*{Appendix}

\label{divisionOfLabor}
\subsection*{Division of labor between the Core library and the plug-ins}
The Core library provides the following functionalities:

\begin{itemize}
\item Drawing of parts that are defined in a XML format
\item Drawing of connections and saving connections to an XML file
\item Selecting, inserting, and moving items on the canvas
\item Undo and redo capabilities
\item Memory management for parts and connections
\item Loading C libraries as separate threads (i.e. parallel processing)
\item Indexing variables in a model, allowing easy look-up
\item Parsing math formulas using muparser library (muparser.sourceforge.net)
\item Command window that plug-ins can use for command-based interaction
\end{itemize}

The plug-ins provide the following features:

\begin{description}
\item[Parts Tree] Provides a window displaying all the parts listed in an XML file.
\item[Connections Tree] Provides a window displaying all the connections listed in an XML file. Each connection is restricted to certain types of parts. For example, a ``transcription regulation" can only be made between a ``transcription factor" and a ``regulator" part.
\item[Part Insertion Tool] Monitors the Parts Tree and allows user to insert parts from the tree.
\item[Connection Insertion Tool] Monitors the Connections Tree and allows user to make permitted connections. 
\item[Selection Tool] Highlights selected items and handles a few other user interface details.
\item[Copy Paste Tool] Provides the copy, cut, and paste feature as well as aliasing and replacing a part's visual representation (see section \ref{standard_visual_format}). 
\item[Name and Family Tool] Displays the family information of items and allows users to change the family of items.
\item[Collision Detector] Detects collision events and reports to other plug-ins. This information is used by several other plug-ins to respond when user places one item adjacent to another.
\item[Container Tool] Allows user to drag items into a container or module (explained in section \ref{modular_design}). Also displays the parameter window.
\item[Plot Tool] Provides the 2D graphing window that can handle multiple graphs. C and Python programs can output to the graph window directly.
\item[Module Tool] Allows construction of modules with interfaces that can be connected (explained in section \ref{modular_design})
\item[Gene Regulatory Network Tool] Provides various automated features that allows easier construction of genetic networks. For example, when a user places a promoter upstream of a gene, the gene's transcription rates are automatically updated. Similarly, when a user connects a transcription factor to a regulator, the kinetics are automatically generated.
\item[Antimony Tool] Uses libAntimony (antimony.sourceforge.net) 
\end{description}

\subsection*{Sample Python scripts}

\textbf{Linear algebra operations on the Stoichiometry matrix}

The following script gets the list of reactions that are selected by the user and generates a stoichiometry matrix for just the selected reactions. It then uses SciPy to perform QR and LU factorization on the stoichiometry matrix. Most of the output is not shown for reading clarity.

\begin{verbatim}
>>A = selectedItems()
>>N = stoichiometry(A)
>>print N[0]

('s1', 's2', 's3', 's4', 's5')

>>print N[1]

('J0', 'J1', 'J2', 'J3', 'J4', 'J5')

>>M = matrix(N[2])
>>print M

[[-1.  0.  0.  0.  1.  0.]
 [ 1. -1.  0.  0.  0.  0.]
 [ 0.  1. -1.  0.  0.  1.]
 [ 0.  0.  1. -1.  0.  0.]
 [ 0.  0.  0.  1. -1. -1.]]

>>from scipy import linalg
>>print linalg.qr(M)
>>print linalg.lu(M)
\end{verbatim}

\textbf{Get all the part sequences in the model as a multi-FASTA file}

The following script will find all the components in the model that belong to the family called ``DNA", which is the root family for promoters, RBS, genes, and other parts that belong on the DNA. Then the script will get the sequence for each part and print the sequence to the screen in FASTA format.

\begin{verbatim}
import pytc
A = pytc.itemsOfFamily('DNA');

if (len(A) > 0):
  names = pytc.getNames(A);
  attribs = ('sequence',);
  seqs = pytc.getAllTextNamed(A,attribs);

  n = len(seqs);
  s = '';
  for i in range(0,n):
    s += '>' + names[i] + '\n' + seqs[i] + '\n';

  print s;
\end{verbatim}

\textbf{Getting annotation and sequence information about a module and its sub-components}

The following script will find a module with a particular name and obtain all of its sub-parts. Then it will loop through the sub-part and print annotation information and sequence for each sub-part. The part which is being accessed in this script is the one shown in figure \ref{fig:composite_part}. Names of people in the output have been replaced with ``no name", and some of the longer sequences have been truncated.

\begin{verbatim}
>>part = find("BBa_I0462")
>>print annotation(part)
('no name', '12/04/2003', 'luxR Protein Generator', 'uri', 'reference')

>>print getTextAttribute(part,"format")
BBa

>>print getTextAttribute(part,"type")
composite

>>subparts = getChildren(part)
>>for i in subparts: print getName(i) + " is a " + getFamily(i);
BBa_I0462_B0012 is a Terminator
BBa_I0462_B0010 is a Terminator
BBa_I0462_C0062 is a Coding
BBa_I0462_B0034 is a RBS

>>for i in subparts: print getName(i) + " : "; print annotation(i);
BBa_I0462_B0012 : 
('no name', 'NA', 'Transcription terminator for the E.coli RNA polymerase.', 'uri', 'reference')

BBa_I0462_B0010 : 
('no name', '11/19/2003', 'Transcriptional terminator consisting of a 64 bp stem-loop', 'uri', 'reference')

BBa_I0462_C0062 : 
('no name', '2003', 'luxR repressor/activator', 'uri', 'reference')

BBa_I0462_B0034 : 
('no name', '2003', 'RBS based on Elowitz repressilator', 'uri', 'reference')

>>for i in subparts: print getName(i) + " : "; print getTextAttribute(i,"sequence");
BBa_I0462_B0012 : 
tcacactggctcaccttcgggtgggcctttctgcgtttata

BBa_I0462_B0010 : 
ccaggcatcaaataaaacgaaaggctcagtcgaaagactgggcctttcgttttatctgttgtttgtcggtgaacgctctc

BBa_I0462_C0062 : 
aaagaggag ... ttctgcgtttata

BBa_I0462_B0034 : 
aaagaggagaaa

>>print getParameter(subparts[3],"strength")
5.0
\end{verbatim}

\pagebreak

{\ifthenelse{\boolean{publ}}{\footnotesize}{\small}
 \bibliographystyle{bmc_article}  
  \bibliography{ref} }     


\ifthenelse{\boolean{publ}}{\end{multicols}}{}

\pagebreak

\section*{Figures}

\subsection*{Figure 1 - Screenshot with a module and a compartment}

\begin{figure}[!h]
\centering
\includegraphics[width=1.0\textwidth]{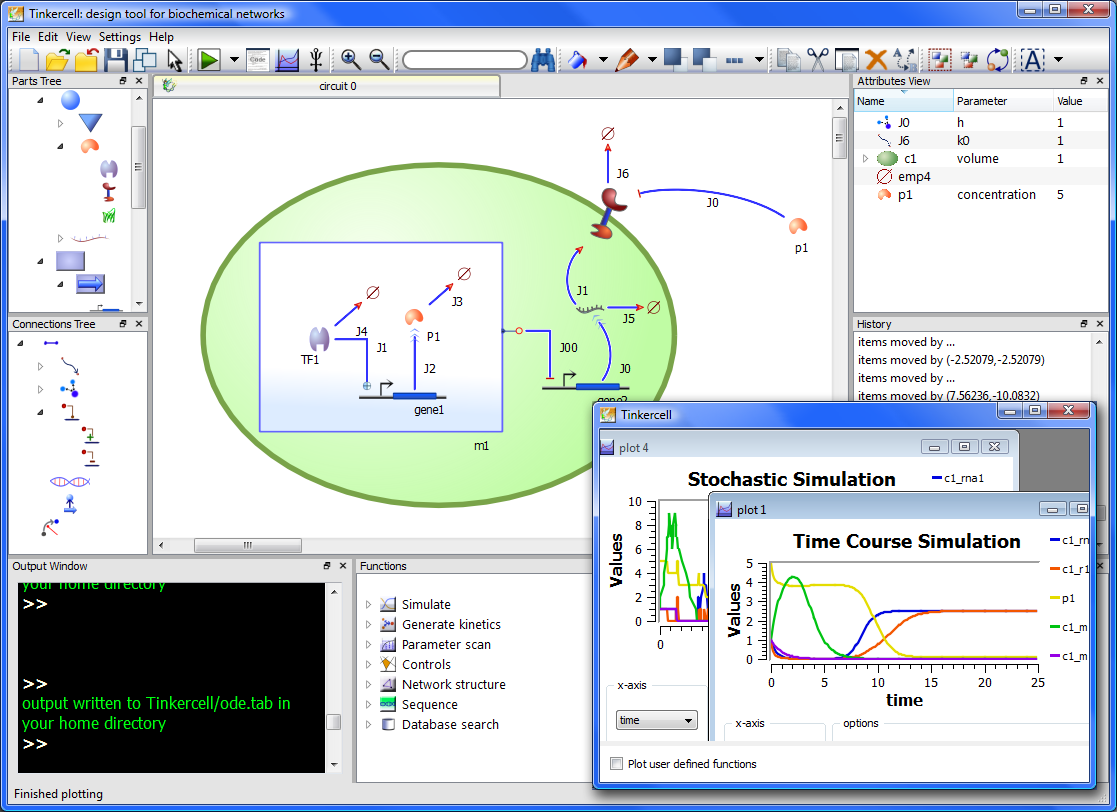}
\caption{This is a screenshot from TinkerCell showing a module inside a compartment. The module (blue rectangle) contains a single gene that is positively regulated. The protein, an ``output" of the module, is the regulator of the gene inside the compartment, which in turn produces a receptor on the compartment membrane.} \label{fig:compartment_module}
\end{figure}

\pagebreak

\subsection*{Figure 2 - Screenshot of a multi-cell system}

\begin{figure}[!h]
\centering
\includegraphics[width=1.0\textwidth]{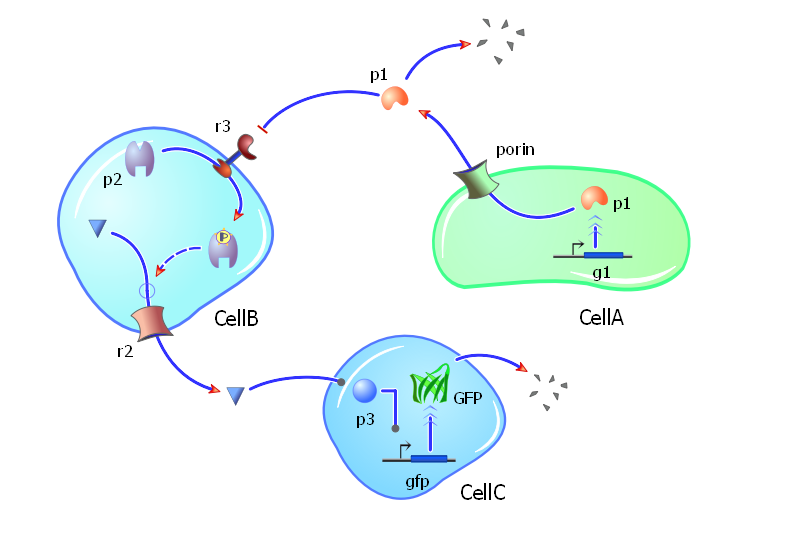}
\caption{This is a screenshot from TinkerCell showing a network involving three cells. Visual models in TinkerCell are very much like cartoon diagrams that are used in textbooks to illustrate pathways. Thus, TinkerCell can serve as an environment for both modeling and illustration. } \label{fig:threecells}
\end{figure}

\pagebreak

\subsection*{Figure 3 - The TinkerCell main window }

\begin{figure}[!h]
  \begin{center}
    \includegraphics[width=1.0\textwidth]{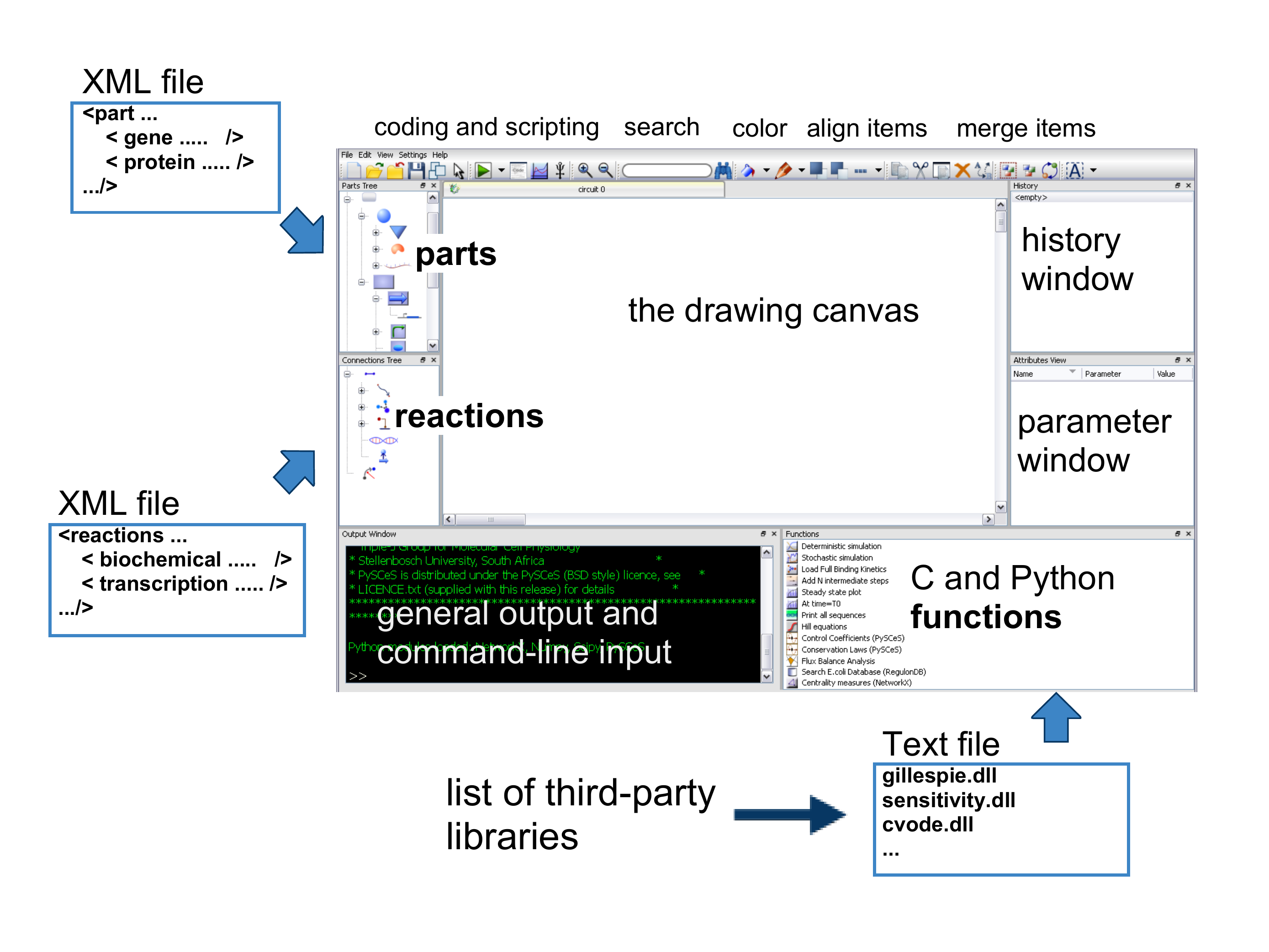}
  \end{center}
  \caption{ Shown above is the initial screen when TinkerCell starts (excluding the labels). The central area is the canvas upon which the network is drawn. The network is constructed using the items in the ``parts" and ``reactions" lists. These lists are loaded from two XML files. In future, these files will represent a standard set of biological parts and reactions. The window at the bottom right lists all the functions available in TinkerCell. These functions are third-party C or Python libraries. The list of functions is loaded from a text file, so new C and Python functions can be added to the list simply by editing this file. The other windows include the history window and parameter window. The parameter window shows the list of items in the network and allows the user to edit the item names and parameters. The command-line window allows users to use Python commands to interrogate or edit the model. }\label{fig:overview_window}
\end{figure}

\pagebreak

\subsection*{Figure 4 - TinkerCell's layered structure}

\begin{figure}[!h]
  \begin{center}
    \includegraphics[width=1.0\textwidth]{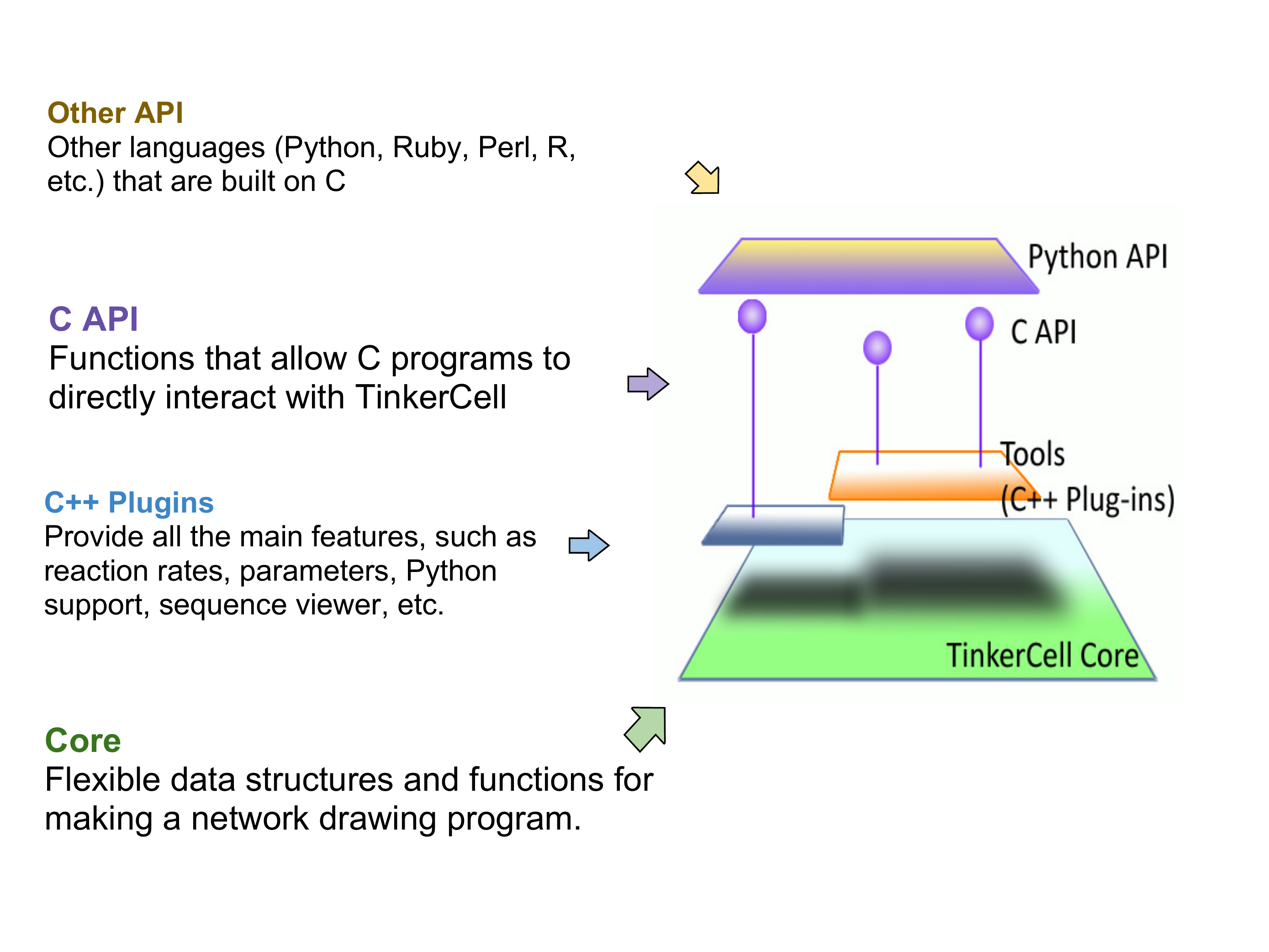}
    \caption{TinkerCell's structure can be described by four layers. The bottom layer, the Core, is a generic library for making network drawing programs. The next layer, plug-ins, are classes written in C++ that perform most of the TinkerCell functions. The plug-ins provide a series of functions that can be called from C programs, and the C function calls can be extended to Python or other scripting languages. }\label{fig:Structure2}
  \end{center}
\end{figure}

\pagebreak

\subsection*{Figure 5 - The Core API}

\begin{figure}[!h]
  \begin{center}
    \includegraphics[width=0.6\textwidth]{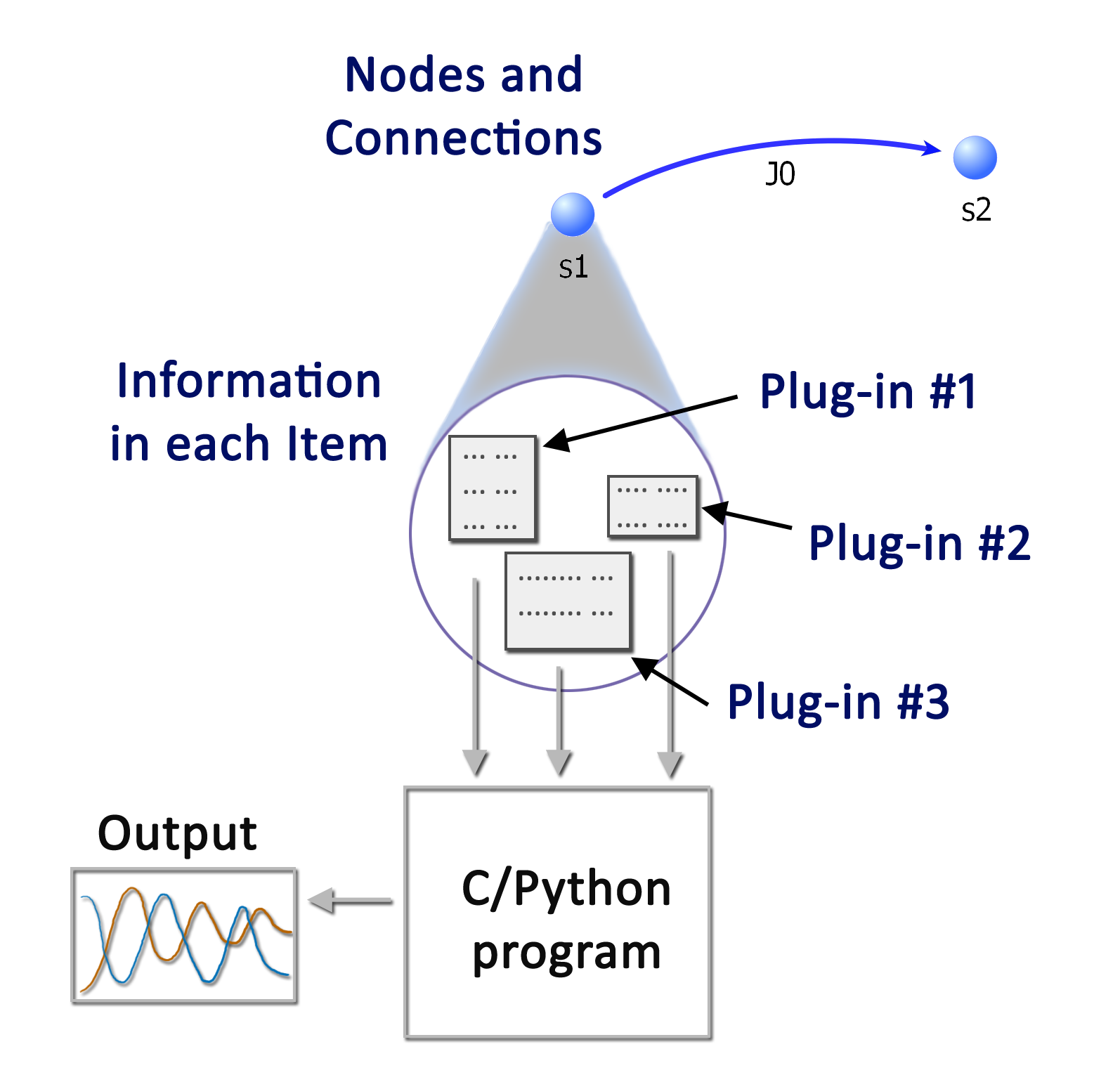}
    \caption{ The structure provided by the Core library. The Core library defines data structures and functions for drawing nodes and connections and storing information inside each node and connection. The various plug-ins utilize this structure and add relevant information to each node and connection. For example, one of the plug-ins adds parameter values to the nodes and connections and another adds reaction rates to the connections. Another plug-in adds position information for cells. This information is utilized by C and Python programs to perform various analyses, such as simulations. Plug-ins also provide graphical interfaces for viewing and editing the information stored inside the items. }\label{fig:Structure1}
  \end{center}
\end{figure}

\pagebreak

\subsection*{Figure 6 - Example plug-ins: reaction rates and parameters}

\begin{figure}[!h]
  \begin{center}
    \includegraphics[width=1.5\textwidth,viewport=0 80 1080 580,clip]{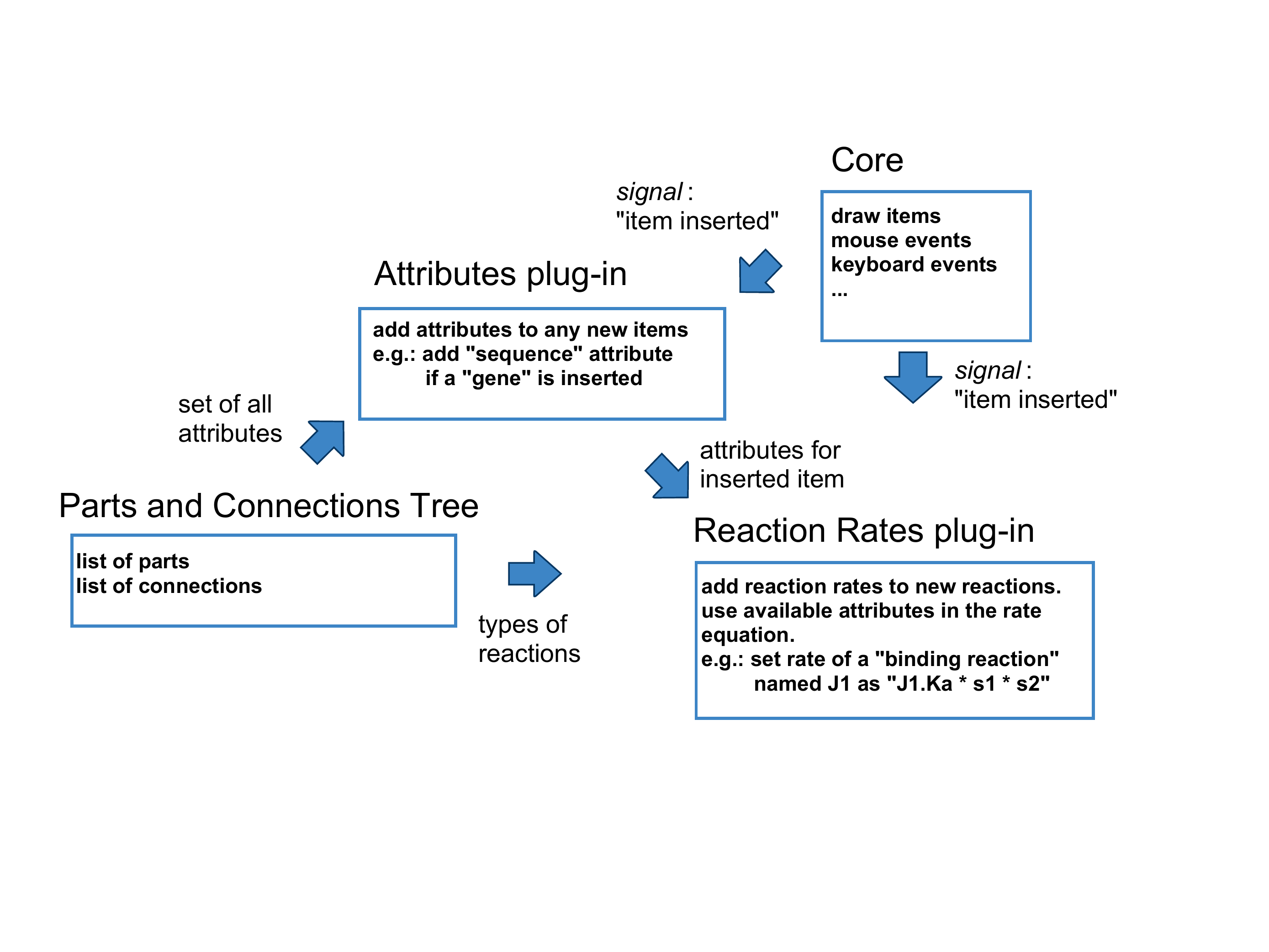}
    \caption{This figure illustrates how two of the plug-ins, the Numerical Attribute plug-in and the Rates plug-in, function together. The Numerical Attributes plug-in is responsible for adding the attributes specified in the tree of parts and connections. The Rates plug-in uses those attributes to generate the default rate equations. The core library informs the plug-ins when a user has inserted a new item. }\label{fig:Attributes_and_rates_example}
  \end{center}
\end{figure}

\pagebreak

\subsection*{Figure 7 - Connecting modules}

\begin{figure}[!h]
   \centering
    \subfloat[ Phosphorylation/dephosphorylation module ]{\label{fig:module_a}\includegraphics[width=0.5\textwidth]{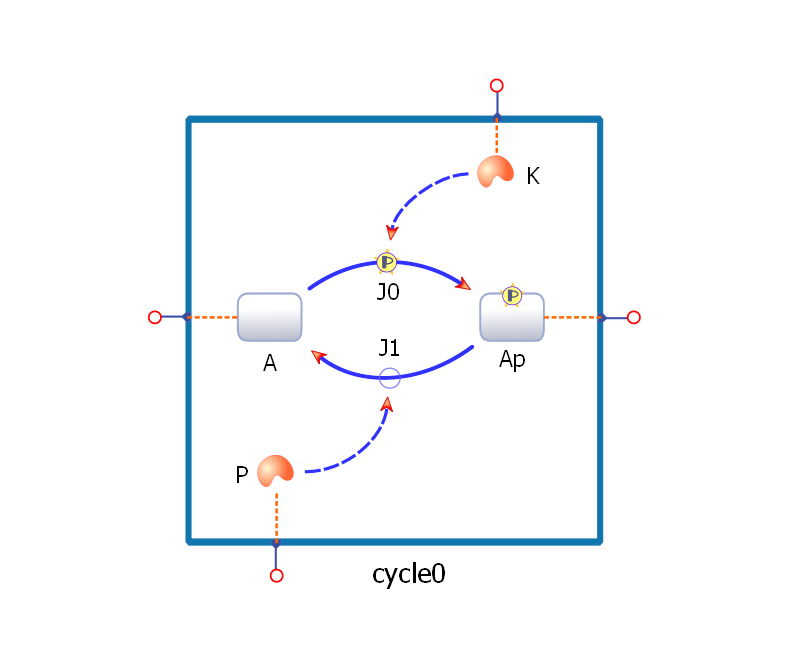}}                
    \subfloat[Three connected modules]{\label{fig:module_b}\includegraphics[width=0.5\textwidth]{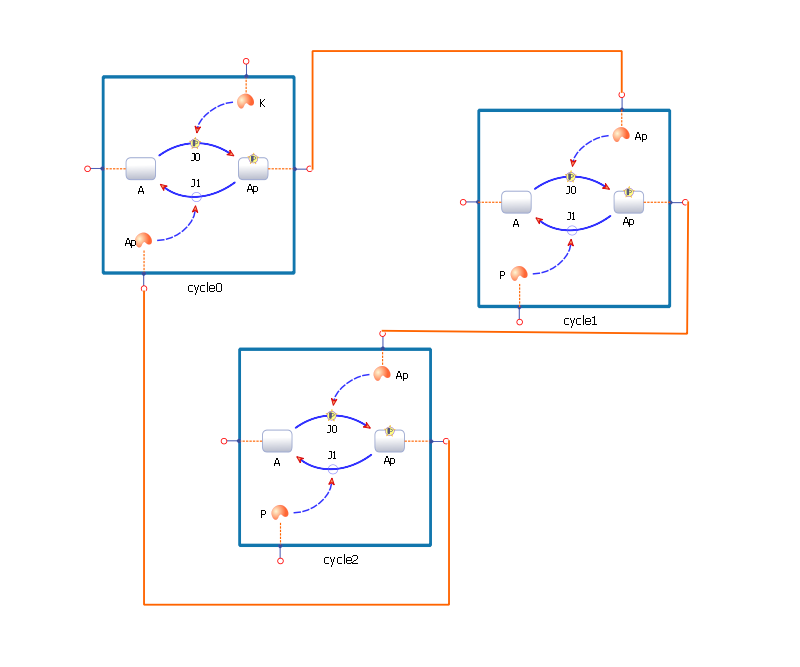}}
  \caption{ (a) A simple phosphorylation/dephosphorylation cycle represented as a module. The user can specify when components in a module should serve interfaces. All four proteins in this module are interfaces, hence there are four pins around each module, one for each of the components in the module. (b) The pins are used to connect one module to another. The module shown in (a) is duplicated three times. The three modules are connected together to form an oscillating network \cite{sauro1993bng}. The orange connections between the module pins indicate that the components that are connected are the same. For example, the phosphorylated protein in the first module is the kinase in the second module, and the phosphorylated protein in the second module is the kinase in the third module. }\label{fig:KinaseModule}
\end{figure}

\pagebreak

\subsection*{Figure 8 - A genetic module}

\begin{figure}[!h]
  \begin{center}
    \includegraphics[width=1.0\textwidth]{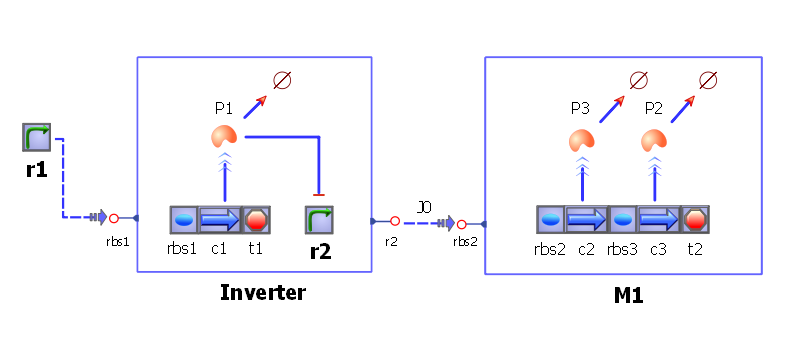}
  \end{center}
  \caption{A simple genetic module as they are commonly used in the synthetic biology community. The first module, the Inverter, is usually placed under the control of some promoter, r1. When r1's activity is high, r2 is low, and when r1 is low, r2 is high, hence the inversion. The connection between promoter r1 and the Inverter module can be thought of as the flux of RNA polymerase. The connection between the Inverter module and the M1 module is also the same. Therefore, r2 controls the transcription rate of the genes in module M1. }\label{fig:PoPS}
\end{figure}

\pagebreak

\subsection*{Figure 9 - BioBrick composite part as a module}

\begin{figure}[!h]
  \begin{center}
    \includegraphics[width=0.5\textwidth]{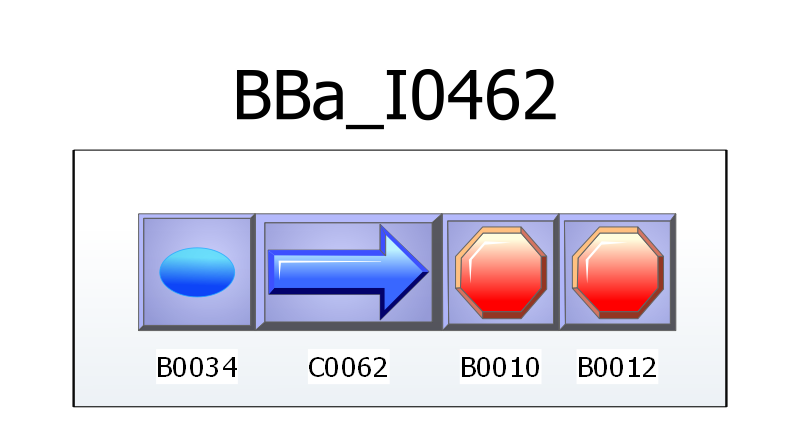}
  \end{center}
  \caption{ A composite part is a biological part composed of other parts, as it is shown here. The name BBa refers to the BioBrick standard that this part conforms to \cite{registryofparts}. Together with the number, BBa I0462 is a unique ID for the composite part. The same naming scheme applies to each of the sub-parts. The composite part itself has its own set of parameters, annotation, authorship, and other information, and the sub-parts retain their individual attributes, such as sequence and parameter values. All this information can be accessed through Python scripts or the user interface. }\label{fig:composite_part}
\end{figure}
\pagebreak

\subsection*{Figure 10 - Ontology of parts and connections}

\begin{figure}[!h]
   \centering
    \subfloat[Part and connection families]{\label{fig:partfamilies}\includegraphics[width=0.7\textwidth]{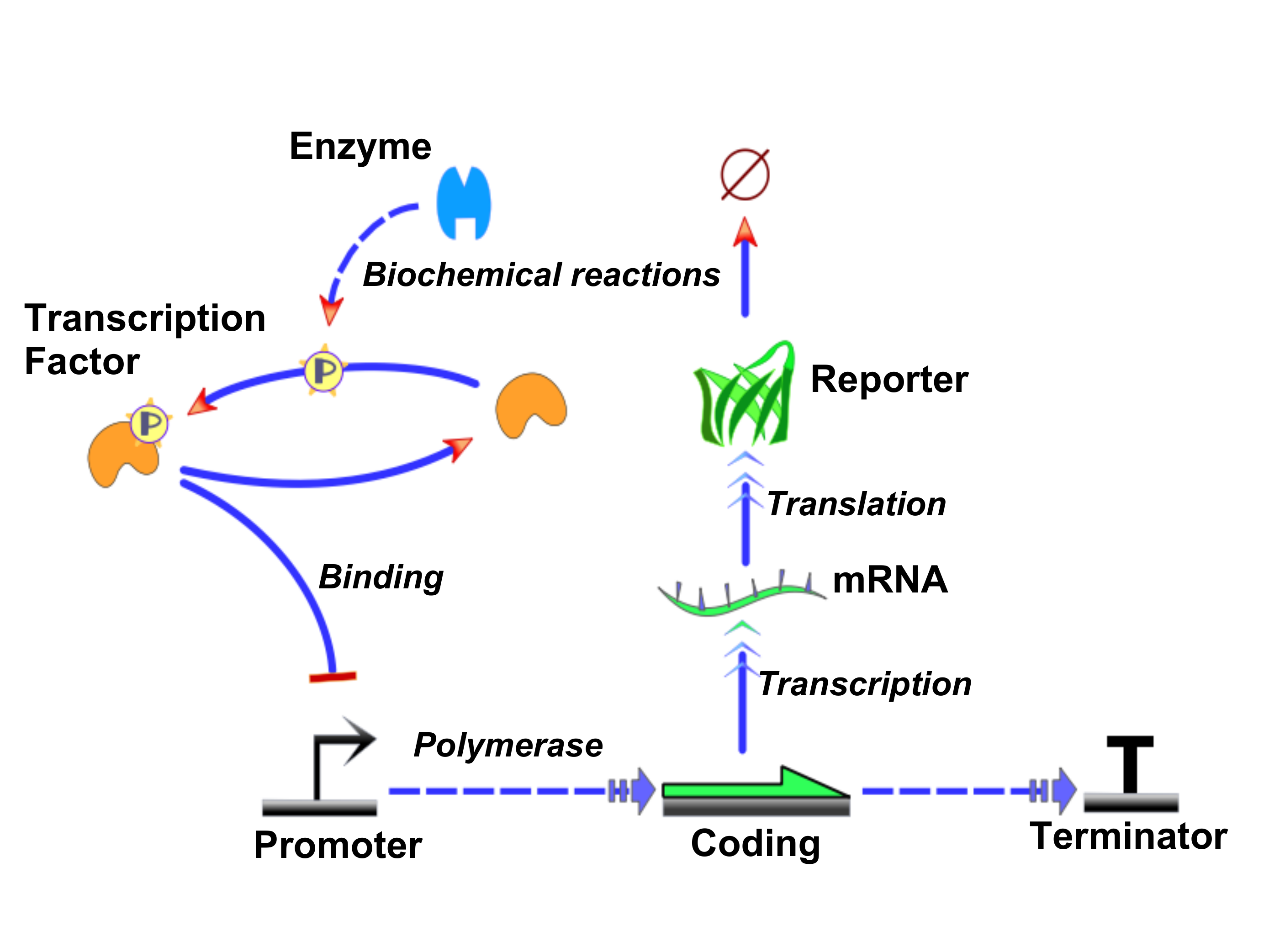}}
	\subfloat[Hierarchy of parts]{\label{fig:partstree}\includegraphics[width=0.2\textwidth]{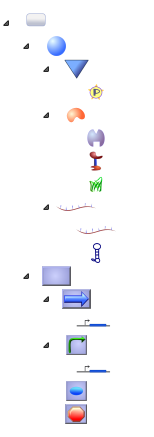}}  
    \caption{Each item in TinkerCell belongs to a family type. A family structure is important if a model should support ontologies or interact with databases. (a) This is a screenshot from Tinkercell with labels indicating the family that each part or connection belongs to. Each connection has a different arrowhead, indicating its family; the arrowheads can be replaced through a visual interface or through a script.(b) This figure shows the current set of part families in TinkerCell, which is loaded from an XML file. The connection families are also loaded in the same way, so the ontology used in TinkerCell can easily be replaced when such a standard is set. The visual appearance of each part and connection arrowheads can be defined in the XML file containing the ontology. }\label{fig:families}

\end{figure}

\pagebreak

\subsection*{Figure 11 - Viewing DNA sequence}

\begin{figure}[!h]
  \begin{center}
    \includegraphics[width=1.0\textwidth]{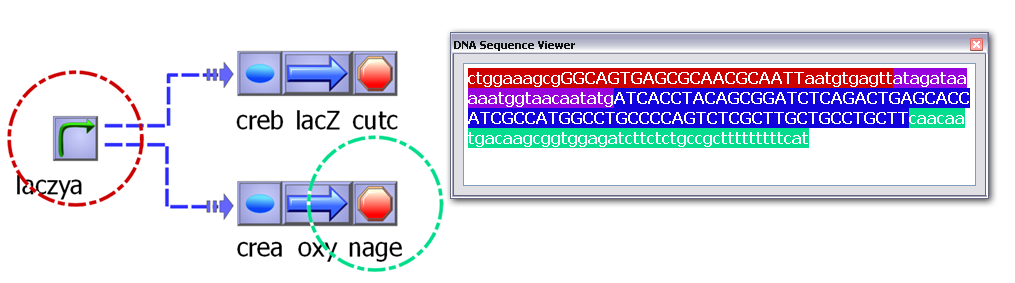}
    \caption{In the screenshot above, several biological parts are connected together. However, the promoter named ``laczya" is regulating two other gene segments. The regulation is shown by the two dotted arrows. Since the promoter is regulating two genes, it is implied that there are two copies of this promoter, one upstream of each gene that it regulates. This is one method of showing regulation in an abstract way in the synthetic biology community \cite{kelly2009ma}. The color coded circles reflect the first and last parts in the displayed sequence. The individual sequences are loaded from RegulonDB \cite{gamacastro2007rvg} via a Python program. The sequence attribute is only available for items of specific families, as described by the family tree of biological parts. The sequence viewing window is provided by a small plug-in. }\label{fig:SequenceView}
  \end{center}
\end{figure}

\pagebreak

\subsection*{Figure 12 - Drawing program for generating graphical objects}

\begin{figure}[!h]
   \centering
    \subfloat[Drawing program for generating graphical objects]{\label{fig:makeParts}\includegraphics[width=0.3\textwidth]{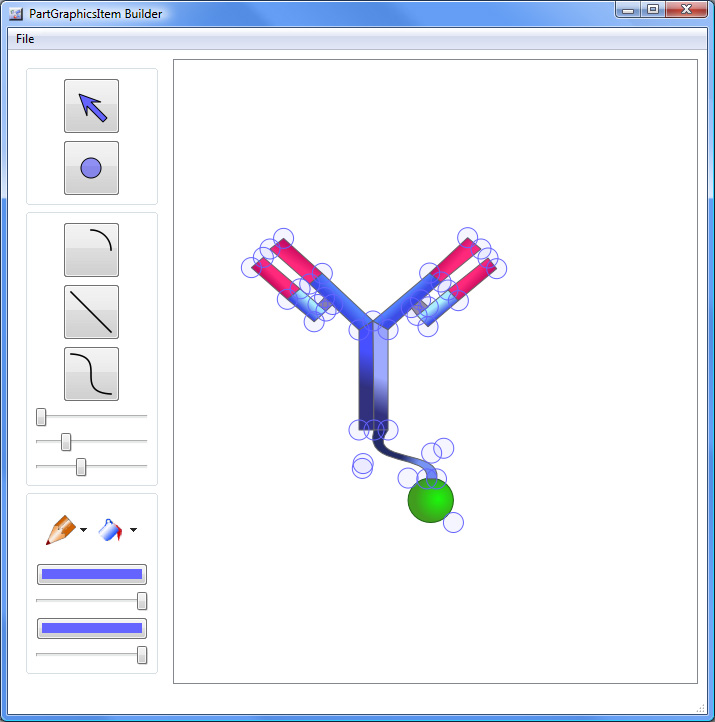}}\ \               
    \subfloat[Replacing the visual representation]{\label{fig:replaceParts}\includegraphics[width=0.25\textwidth]{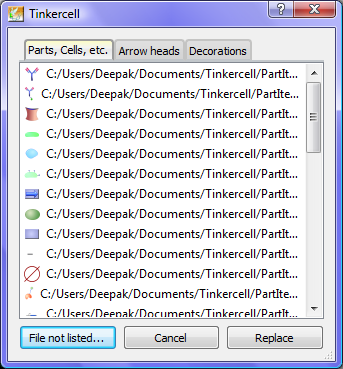}}
  \caption{(a) TinkerCell comes with various visual representations for various types of parts, such as genes, promoters, proteins, transcription factors, receptors, fluorescent reporters, and cells. These visual objects are created by a polygon drawing program available with TinkerCell. (b) Visual representations of objects, including arrow heads or decorators, in TinkerCell can be replaced using the ``replace graphics" dialog.}\label{fig:replace_graphics}
\end{figure}

\pagebreak

\subsection*{Figure 13 - Simple way of modeling gene regulatory networks}

\begin{figure}[!h]
  \begin{center}
    \includegraphics[width=1.0\textwidth]{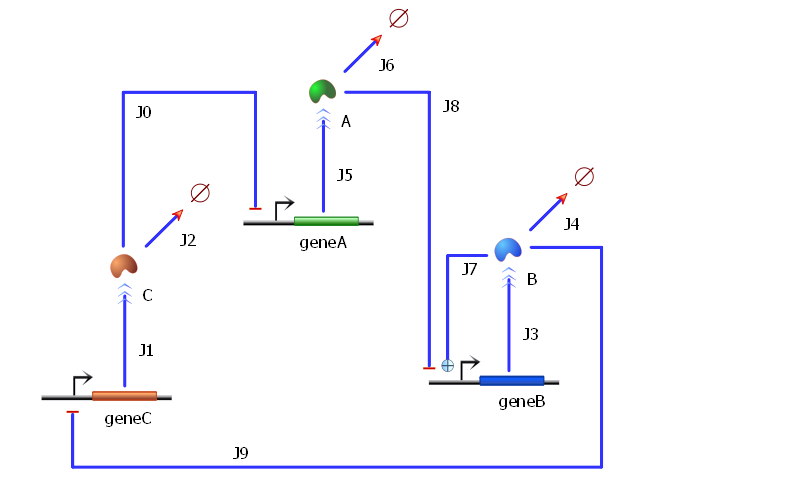}
    \caption{Modeling gene regulatory networks using fractional saturation models where the transcription factor association and dissociation from the operator site are assumed to be in equilibrium \cite{shea1985oc}. Under this assumption, the protein production rate can be modeled as being proportional to the ratio of the active promoter states to all possible promoter states. }\label{fig:grn_simple}
  \end{center}
\end{figure}

\pagebreak

\subsection*{Figure 14 - Gene regulatory networks using parts}

\begin{figure}[!h]
  \begin{center}
    \includegraphics[width=1.0\textwidth]{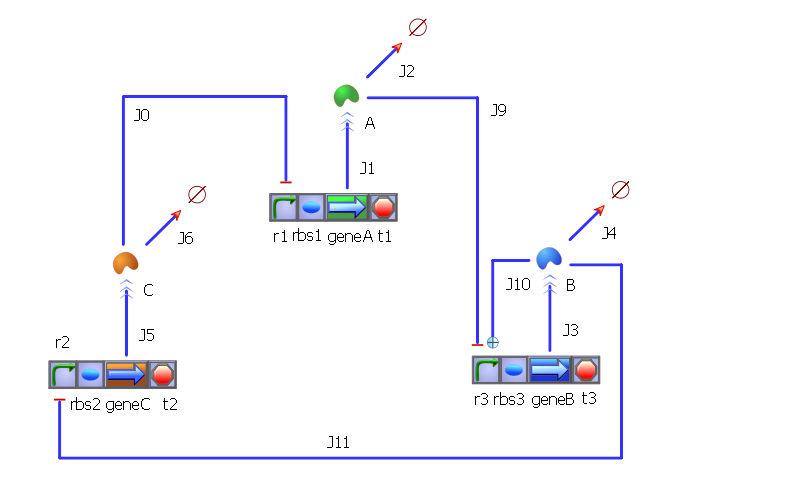}
    \caption{Modeling gene regulatory networks using separate parts. This method of constructing genetic networks also uses fractional saturation models, similar to Figure \ref{fig:grn_simple}. In constrast to Figure \ref{fig:grn_simple}, there is more flexibility in the rate expressions. For example, when a user changes the promoter located upstream of a coding region, the rate expression for transcription is automatically adjusted to take into account the strength parameter of the new promoter. If translation is modeled explicitly, changing the RBS will have a similar effect. This method of constructing genetic networks is preferred when the user intends to swap promoters or RBS without having to change the rate equations explicitly. }\label{fig:grn_parts}
  \end{center}
\end{figure}

\pagebreak

\subsection*{Figure 15 - Gene regulatory network using delays and without using equilibrium assumptions}

\begin{figure}[!h]
  \begin{center}
    \includegraphics[width=1.0\textwidth]{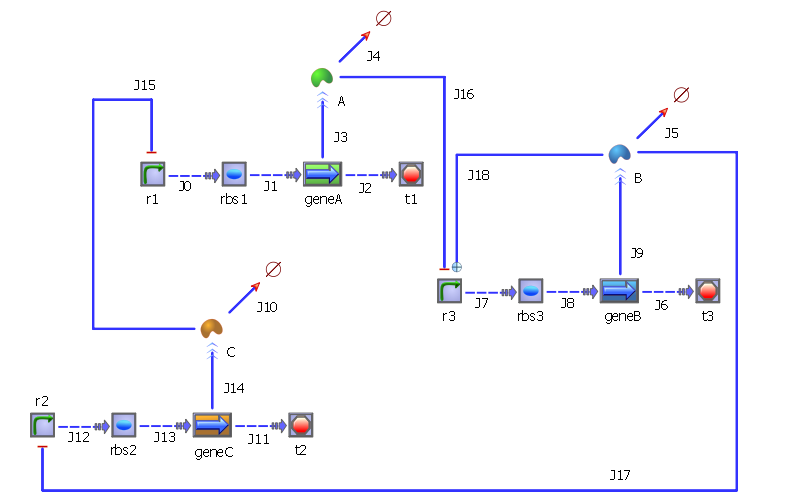}
    \caption{Modeling gene regulatory networks without using equilibrium assumptions. The method of modeling represents the transcription and translation process as multiple reactions. These reactions represent the polymerization process catalyzed by RNA polymerase and the ribosomes. Because the process is modeled as separate reactions, there will be a delay for the entire process to complete. Transcription factor binding and unbinding can be modeled explicitly; since these are single reactions, they will be much faster than the transcription and translation process. While this technique can be thought of as an alternative to models that use time delays, the technique also allows modeling events such as RNA polymerase dissociating from the DNA in the middle of transcription. }\label{fig:grn_full}
  \end{center}
\end{figure}

\pagebreak

\subsection*{Figure 16 - User interface via C programs}

\begin{figure}[!h]
  \begin{center}
    \includegraphics[width=0.35\textwidth]{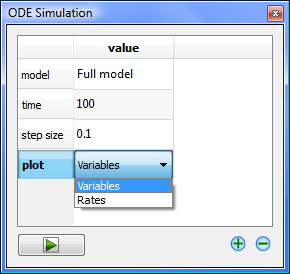}
  \end{center}
  \vspace{-10pt}
  \caption{This window is created from a C program using a function called ``createInputWindow" that is provided by the C API. The Python API also has the same function. The input from this window is converted to a matrix and handed directly to the calling C or Python program. A handful or such functions are available that allow C or Python programs to interact with the user.}\label{fig:createInputWindow}  
\end{figure}

\pagebreak

\subsection*{Figure 17 - Automatically generating intermediate complexes}

\begin{figure}[!h]
   \centering
    \subfloat[Binding reactions]{\label{fig:binding_a}\includegraphics[width=0.3\textwidth]{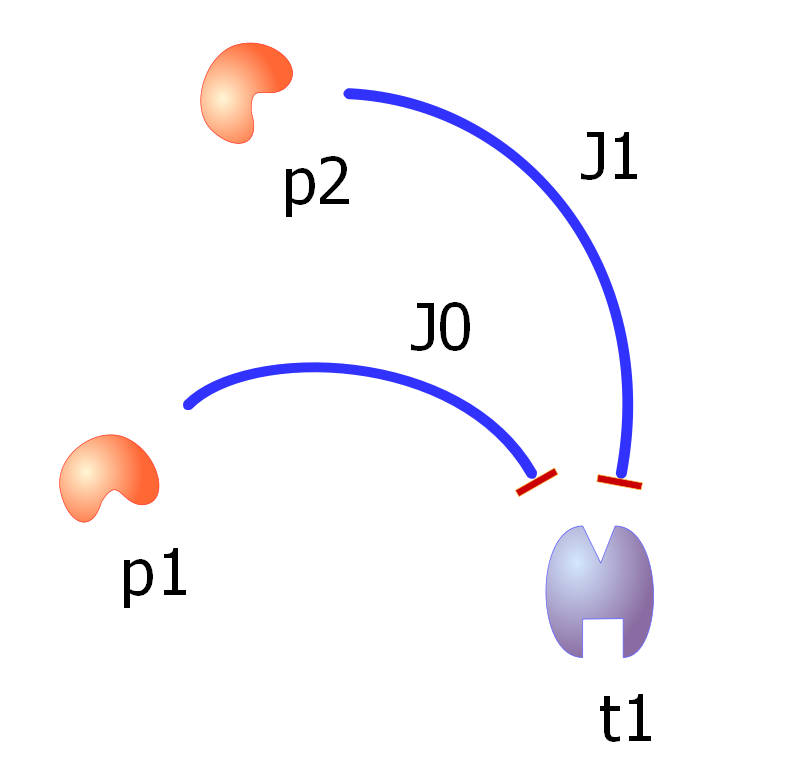}}                
    \subfloat[Complete set of binding reactions]{\label{fig:binding_b}\includegraphics[width=0.5\textwidth]{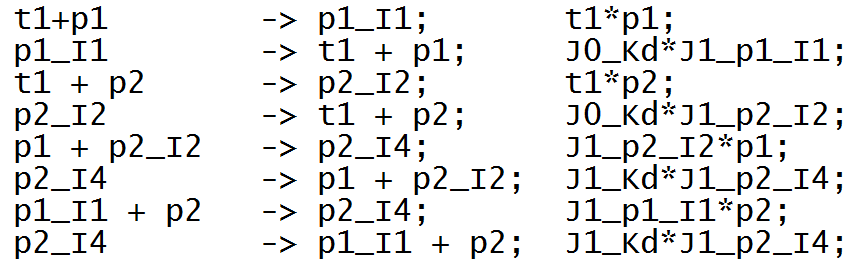}}
  \caption{ (a) This screenshot represents two proteins, p1 and p2, that bind t1. There are six different molecular species is in this network: p1, p2, t1, p1+t1, p2+t1, p1+p2+t2. By default, the original form, i.e. t1, is the active protein, which is why the arrow-heads on the reaction arcs in the (a) represent inhibition. The number of different species will increase exponentially as the number of individual molecules increase. Modeling all the details cannot be achieved without some form of automation. (b) Listed here are all the binding and unbinding kinetics that are automatically generated in TinkerCell. The names with an I imply ``intermediate". These reactions are generated by a small C program called ``Load Full Binding Kinetics", which can be accessed through TinkerCell's visual interface. The visual display remains clean because the reactions are embedded inside each reaction arc, i.e. each reaction drawn on the screen represent more than one reaction. }\label{fig:binding_kinetics}
\end{figure}

\pagebreak

\subsection*{Figure 18 - Automatically generate intermediate steps}

\begin{figure}[!h]
   \centering
    \subfloat[A slow process]{\label{fig:tabasco_a}\includegraphics[width=0.4\textwidth]{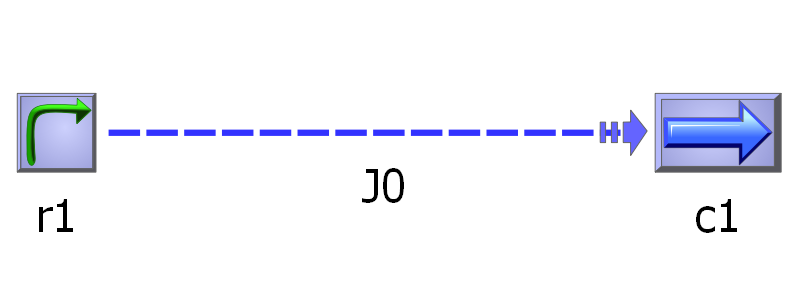}}                
    \subfloat[Complete set of reactions]{\label{fig:tabasco_b}\includegraphics[width=0.4\textwidth]{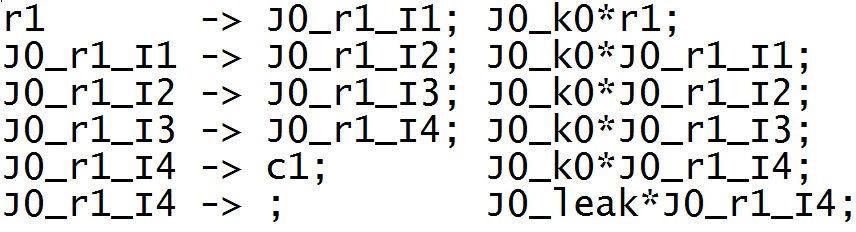}}
 \caption{ (a) This screenshot represents a different way of modeling gene regulation by modeling the delay in transcription using multiple steps \cite{kosuri2007ts}. In this method, the movement of RNA polymerase through the DNA is modelled explicitly. This is achieved by adding N (user's choice) intermediate steps between the r1 and c1, where r1 represents ``active" promoter, or RNA polymerase bound to r1, and c1 represents ``active" gene. There is a delay between the time RNA polymerase binds the promoter and transcribes the gene. (b) Listed here are the set of reactions that is represented by the diagram shown in (a). The names with an I imply ``intermediate". The number of intermediate steps is five, but the C program that generates the set of reactions asks the user to enter this number. }\label{fig:tabasco_kinetics}
\end{figure}

\pagebreak

\subsection*{Figure 19 - Graphical output from C/Python programs}

\begin{figure}[!h]
\begin{center}
    \includegraphics[width=0.9\textwidth]{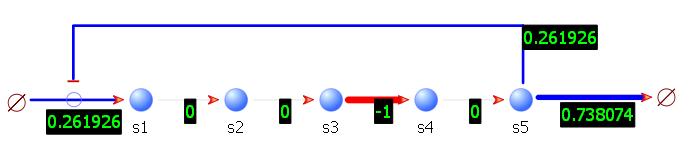}
  \end{center}
  \caption{ This screenshot is the output from a Python script that generates the PySCeS model using the TinkerCell Python API. The script uses PySCeS \cite{olivier2005mcs} to compute the control coefficients and adjusts the line widths in the network according to the output from PySCeS. Negative control coefficients are colored red and positive ones are colored blue. A similar visual output is produced by functions that perform calculations on the network, such as flux balance analysis. }\label{fig:Visual_C_outputs}
\end{figure}

\pagebreak

\subsection*{Figure 20 - Accessing RegulonDB}

\begin{figure}[!h]
  \begin{center}
    \includegraphics[width=0.5\textwidth]{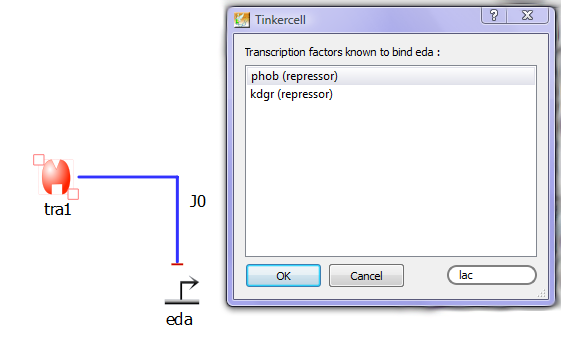}
  \end{center}
  \caption{This screenshot shows the interface provided by the Python script that searches RegulonDB and provides the user with a list of parts from \textit{E.coli}. The script uses the selected part's family information to identify it as a promoter, RBS, coding sequence, or transcription factor. It then uses this information to search the appropriate types of parts in RegulonDB. Additionally, the script looks at the connections made by the part to prune the list. For example, if a transcription factor is regulating the selected promoter, then the Python script will only show the list of binding sites that are targeted by the particular transcription factor.}\label{fig:regulonLoad}  
\end{figure}

\end{bmcformat}
\end{document}